\documentclass{birkmult}
\usepackage{amsfonts,amssymb,amsmath, euscript}
\begin{document}

\title{\bf $U(1)$ invariant Membranes and Singularities}

\author {Jens~Hoppe}
\address{%
Department of Mathematics, Royal Institute of Technology\\
10044, Stockholm, Sweden}                                 

\thanks{Based on talks given at the Banff Centre (July 2007)  and the Institut  Henri Poincare (March 2008)}

\begin{abstract}
A formulation of  $U(1)$ - symmetric  classical membrane  motions (preserving  one  rotational symmetry) is  given, 
and reductions to systems of ODE's, as well as some ideas concerning singularities and integrability.
\end{abstract}

\maketitle

Let me  start by giving  some  particular example(s) of  a  $2$-dimensonal (time-like, time periodic) extremal 
manifold $\mathbb{M}_{2} \subset \mathbb{R}^{1,2}$ and  $3$-folds (singularities included) of  vanishing mean 
curvature, in 
$4$-dimensional Minkowski space (\cite{H1},\cite{H2}):
\begin{equation*}
\mathbb{M}_{2} := \{\!
\left(\!\!
\begin{array}{ccc} 
t \\ x \\ y
\end{array}
\!\!
\right)
\!\in \mathbb{R}^{1,2} |\,\, x\cos 4t - y\sin 4t = 8(x^2+y^2)^{2} - 3(x^2+y^{2})- \frac{3}{32}; \,
x^2+y^{2} > \frac{1}{16}\, \},  
\end{equation*}
\begin{equation}\label{J1}
{\mathbb{M}}^{(t)}_{3}: = \{ (x^{\mu})
\in \mathbb{R}^{1,3}  |\,\, \mathcal{P}(x)\mathcal{P}(y)\mathcal{P}(t)= \mathcal{P}(z)\,\}, 
\end{equation}
\begin{equation*}
{\mathbb{M}}^{(s)}_{3}: = \{ (x^{\mu})
\in \mathbb{R}^{1,3} |\,\, \mathcal{P}(x)\mathcal{P}(y)\mathcal{P}(z)= \mathcal{P}(t)\}, 
\end{equation*}
where  $\mathcal{P}: \mathbb{R} \rightarrow \mathbb{R}_{\geq1}$ is a ($2\omega$ real periodic) elliptic 
Weierstrass function satisfying 
$$
 \mathcal{P}'^2= 4\mathcal{P} (\mathcal{P}^2 - 1), \quad  \mathcal{P}(\omega)=1.
$$
${\mathbb{M}}^{(s)}_{3}$ can be  thought  of  as  a (infinite assembly of) melting  ice-block(s), 
as  at $t=0$ it covers the surface(s) of infinitely many cubes (each of  size $2\omega$), packed  together , 
for  $0<t<T=2\omega$ instantaneously separating into infinitely many melting (round/ed) ice piece(-surface)s,
each of  which shrinks to a point $(t=T/2=\omega)$ and then growing  again  to become , at $T=2\omega$, the 
surface  of infinitely  many ice-cubes, stacked together. $\overline{\mathbb{M}}_{2}$ (with the $x^2+y^2=\frac{1}{16}$
 line of singularities included) gives the  motion of a closed string of  a fixed (heart-like) shape (with exactly one
 cusp - singularity, that arises when  the radius  $r$  takes its minimum, $r=\frac{1}{4}$), rotating with  
constant  angular  velocity (=4)  around  the origin [6].  Before discussing time-like  extremal  3-manifolds  with one  rotational 
 symmetry  embedded in $\mathbb{R}^{1,4}$,  consider for a moment  the general (parametric) description  of 
$(M+1)$-dimensional extended manifolds $\mathbb{M}$ embedded in $D$-dimesional  Minkowski-space: varying the  
volume-functional 
\begin{eqnarray}\label{J2}
{\rm Vol}(\mathbb{M})=\int d^{M+1}\varphi\sqrt{G}= S[x^{\mu}]
\end{eqnarray}
with respect  to the  embedding functions $x^{\mu}(\varphi^{0},...,\varphi^{M}),\,  \mu= 0,..., D-1$, 
one obtains  the parametic  extremal volume  equations, 
\begin{eqnarray}\label{J3}
\frac{1}{\sqrt{G}}\partial_{\alpha}(\sqrt{G}G^{\alpha\beta}\partial_{\beta}x^{\mu})=0,\quad (\mu=0,...,D-1),
\end{eqnarray}
where
$ 
G_{\alpha\beta}:=\partial_{\alpha}x^{\mu}\partial_{\beta}x^{\nu}\eta_{\mu\nu}, \;
\eta_{\mu\nu}={\rm diag}(1,-1,...,-1), \;
G^{\alpha\beta}G_{\beta\gamma}=\delta^{\alpha}_{\gamma},\;
G=(-1)^{M}\det (G_{\alpha\beta}),\;  \alpha,\beta=0,...,M.
$ 

In the gauge $\varphi^{0}=x^{0}(=:t)$,  i.e. 
$(x^{\mu})^T = (t, \vec{x}(t, \varphi^{1},...,\varphi^{M}))$, 
and also demanding
\begin{eqnarray}\label{J4}
\sum_{i=1}^{D-1}\dot{x}_{i}\frac{\partial x_{i}}{\partial\varphi^{r}}=0,\quad (r=1,...,M)\;,
\end{eqnarray}
eq. (\ref{J3})$_{\mu=0}$ becomes  a conservation law 
\begin{eqnarray}\label{J5}
\dot\rho=0,\quad  \rho := \sqrt{\frac{g}{1 - \dot{\vec{x}}^2 }}
\end{eqnarray}
while the remaining ones,   (\ref{J3})$_{\mu=i}$,   become the evolution equations for  a M-dimensional
  time-dependent surface $\Sigma(t)$ moving in $\mathbb{R}^{D-1}$,
\begin{eqnarray}\label{J6}
\ddot{x}_{i}=\frac{1}{\rho}\partial_{r}(\frac{g}{\rho}g^{rs}\partial_{s}x_{i}) \quad (i=1,...,D-1);
\end{eqnarray}
$g^{rs}$  respectively $g$ are the inverse, respectively the determinant, of the  $M$-dimensional metric 
on $\Sigma(t)$,  $g_{rs}$, induced  from $\mathbb{R}^{D-1}$. In  \cite{H3}, I also considered the equations
\begin{eqnarray}\label{J7}
\ddot{\vec{x}}=(1 - \dot{\vec{x}}^2) g^{rs}\partial^2_{rs}\vec{x},
\end{eqnarray}
respectively 
\begin{eqnarray}\label{J8}
\ddot{\vec{x}} &=\frac{1}{\rho^2}gg^{rs}\partial^2_{rs}\vec{x}.
\end{eqnarray}
 The folllowing calculations show that (\ref{J8}) and (\ref{J4}) consistently imply (\ref{J5}) and (\ref{J6}):
\begin{equation}\label{J9}
\begin{split}
\frac{1}{2}(1 - \dot{\vec{x}}^2 )\frac{d}{dt}(\rho^2) &=\frac{1}{2}\dot{g}+g\frac{\dot{\vec{x}}\ddot{\vec{x}}}
{1 - \dot{\vec{x}}^2}\\
&= \frac{1}{2}gg^{rs}\dot{g}_{rs} +\frac{g}
{\rho^2(1 - \dot{\vec{x}}^2)}(gg^{rs}\partial^2_{rs}\vec{x})\dot{\vec{x}} \\
&=gg^{rs}\partial_{r}\vec{x}
\partial_{s} \dot{\vec{x}}+ gg^{rs}\dot{\vec{x}}\partial^2_{rs}\vec{x}=0.
\end{split}
\end{equation}
In order  to deduce that (\ref{J8}) implies (\ref{J6}) one may equivalently show  that ( in a parametrization which 
(\ref{J8}) and (\ref{J4}) hold) necessarily
\begin{equation}\label{J10}
\partial_{r}(\frac{gg^{rs}}{\rho})\partial_{s}\vec{x}=0 \; ;
\end{equation}
multiplying by $\frac{\rho}{g}\partial_{a}\vec{x}$ one obtains 
\begin{equation}\label{J11}
\begin{split}
\frac{\rho}{g}\partial_{a}(\frac{g}{\rho}) + \partial_{r}(g^{rs})g_{sa}
 &=\frac{1}{2}\frac{\partial_{a}g}{g}- \frac{\dot{\vec{x}}  \partial_{a} \dot{\vec{x}}}{1 - \dot{\vec{x}}^2} -
g^{rs}( \partial^2_{rs}\vec{x}\partial_{a}\vec{x} + \partial_{s}\vec{x}  \partial^2_{ra}\vec{x} )\\
 &=\frac{\ddot{\vec{x}}\partial_{a}\vec{x}}{1 - \dot{\vec{x}}^2} - g^{rs}\partial^2_{rs} \partial_{a}\vec{x}=0.
\end{split}
\end{equation}
Let us  now restrict to the case  of membranes ($M=2$), as well as to $\Sigma(t)$ being rotationally symmetric, of 
 the form 
\begin{equation}\label{J12}
\vec{x}(t,\varphi^{1} \varphi^{2})=
\begin{pmatrix}
u_{1}(t,\varphi^{1}) \begin{pmatrix}\cos\varphi^{2}&\\ \sin\varphi^{2}\\\end{pmatrix}
\\
.\\
.\\
.\\
u_{N}(t,\varphi^{1})\begin{pmatrix}\cos\varphi^{2}&\\ \sin\varphi^{2}\\ \end{pmatrix}\\
\end{pmatrix}.
\end{equation}
Noting that $(g_{rs})$, hence also  $g, g^{rs}, $ and  $\rho$, will be  independent of $\varphi^{2}$,
$\vec{e}(t,\varphi):=a\vec{u} (\frac{t}{a},\varphi^{1}(\varphi))$, with $\varphi(\varphi^{1})$, respectively 
$\varphi^{1}(\varphi)$ defined  via 
$a^2\frac{d\varphi}{d\varphi^{1}}=\rho(\varphi^{1})$ (where $a^2:=<\rho>$, the average value of $\rho$)
equation(s) (\ref{J6}) may be written in the form
\begin{eqnarray}\label{J13}
\ddot{\vec{e}}=\vec{e}^{\,\,2}\, \,
{\vec{e}}^{\,''}+ 2(\vec{e}\,\,\acute{\vec{e}})\,\acute{\vec{e}}
-\acute{\vec{e}}^{\,\,2}\,\vec{e}\;,
\end{eqnarray}
to be supplemented  by the two   constraints (cp. (\ref{J4})/(\ref{J5}))
\begin{eqnarray}\label{J14}
\dot{\vec{e}}\,\acute{\vec{e}}=0,
\end{eqnarray}
\begin{eqnarray}\label{J15}
\dot{\vec{e}}^{\,\,2} + \vec{e}^{\,\,2}\acute{\vec{e}}^{\,\,2}=1.
\end{eqnarray}
Note that, when $D=5 \, (N=2)$, and  as long as $\dot{\vec{e}}$ and $\acute{\vec{e}}$ are linear independent, these 2 conditions 
 (\ref{J14}) and  (\ref{J15})  actually imply (\ref{J13}), - which can be  seen by comparing 
the 4 equations  obtained 
 from  differentiating (\ref{J14}) and  (\ref{J15}) with respect to $t$ and $\varphi$  with the scalar product of  (\ref{J13})
with $\dot{\vec{e}}$, respectively  $\acute{\vec{e}}$.
Also note that (\ref{J14}) and  (\ref{J15}), which may be thought of as evolution equations of a  string
  moving in the plane, $\dot{\vec{e}}=v\vec{n}$ (with $\vec{n}$ being  the unit normal to $\acute{\vec{e}}$ 
and the magnitude of the normal velocity  being $\sqrt{1-\vec{e}^{\,\,2}\acute{\vec{e}}^{\,\,2}})$, consistently
 reduce  to a system of ODE's when making  the Ansatz
\begin{equation}\label{J16}
\vec{e}(t,\varphi)=R(\omega t)\vec{f}(\varphi+ c\,t),
\end{equation}
where $R(\omega t)$ is given by 
$$
R(\omega t)=e^{\begin{pmatrix}0&1\\-1&0\\
\end{pmatrix}\omega t}=\begin{pmatrix}
\cos{\omega t}&\sin{\omega t}\\
-\sin{\omega t}&\cos{\omega t}\\
\end{pmatrix}
\in SO(2)
$$
and the possible shapes $\vec{f}=r\left( \begin{array}{c}
                        \cos \theta      \\
                        \sin \theta   \\
                              \end{array} \right)$ being determined by
solving
\[
\omega^2 (u-c^2)^2 {u'}^2=4(\omega^2u-1)(c^2-\omega^2u^2)
\]
in terms of elliptic integrals involving
$u=r^2(\tilde{\varphi}:=\varphi+ct)$, and $\frac{d \theta}{d
\tilde{\varphi}}=\frac{\omega^2 u -1}{c^2-u}\frac{c}{\omega u}$

A more general approach to solving (\ref{J14}) and  (\ref{J15}) results when writing $\vec{e}$ in polar form, 
\begin{equation}\label{J17}
\vec{e}(t,\varphi)=r(t,\varphi) \begin{pmatrix}
\cos\psi(t,\varphi)\\
\sin\psi(t,\varphi)\\
\end{pmatrix}\;,
\end{equation}
and  the corresponding equations,
\begin{equation}\label{J18}
\begin{split}
\dot{r}\acute{r}+r^2\dot{\psi}\acute{\psi}&=0,\\
\dot{r}^2 +{r}^2\dot{\psi}^{2} + {r}^2({\acute{r}}^2+& {r}^2\acute{\psi}^{2})=1
\end{split}
\end{equation}
as 
\begin{equation}\label{J19}
(r\dot{\psi}\pm {r}^2\acute{\psi})^2 + (\dot{r}\,\pm r\acute{r})^2=1.
\end{equation}
Eq. (\ref{J13}), on the other hand, is consistent with the Ansatz (cp. \cite{H4} for  the $N=1$ case)
\begin{equation}\label{J20}
\vec{e}=t^{a}\vec{f}(t^{b}\varphi),\quad \,\, (a+b+1=0)\;,
\end{equation}
yielding  the system of ODE's
\begin{equation}\label{J21}
\vec{f}^2{\vec{f}}^{\,\,''} + 2(\vec{f}\,\acute{\vec{f}})\acute{\vec{f}}- \acute{\vec{f}}^2
=a(a-1)\vec{f} +(2-a)(a+1)s\vec{f}+ (a+1)^{2}s^2{\vec{f}}^{\,\,''}
\end{equation}
for $\vec{f}(s)$. For $a=-1$, eq.  (\ref{J21}) follows from 
\begin{equation}\label{J22}
H= \frac{1}{2}\,\frac{\vec{p}^{\,\,2}}{\vec{f}^{\,\,2}}- \vec{f}^{\,\,2} \quad   \hbox{resp.}\quad
L= \frac{1}{2}\vec{f}^{\,\,2} {\acute{\vec{f}}}^{\,\,2} +\vec{f}^{\,\,2}.
\end{equation}
While  the system separates  for  general  $N$, a particularly simple solution can be obtained   for  $N=2$, for  which
\begin{equation}\label{J23}
L= \frac{1}{2}\, r^{2}(\acute{r}^{\,\,2} + r^{2}\acute{\theta}^{\,2}  + 2)
\end{equation}
when using $\vec{f}^{\,\,T}= r(s)\,(\cos{\theta}(s),\, \sin{\theta}(s))$.
\begin{equation}\label{J24}
\frac{d}{ds}(r^{4}\,\acute{\theta})=0,\quad \acute{\theta}=\frac{L}{r^{4}}
\end{equation}
allows to eliminate $\acute{\theta}$,
\begin{equation}\label{J25}
L= \frac{1}{2}\, r^{2}\,(\acute{r}^{\,\,2} +  \frac{1}{2}\,\frac{L^2}{r^{4}}) + r^{2}\;,
\end{equation}
and 
\begin{equation}\label{J26}
\frac{1}{2}\, r^{2}\,\acute{r}^{\,\,2} +  \frac{1}{2}\,\frac{L^2}{r^{4}} - r^{2}= E
\end{equation}
to be constant, leading to the elliptic integral, also obtained by  M.~Trzetrzelewski 
(private communication), for $v= r^{2}$
\begin{equation}\label{J27}
\int ds =\pm\int\frac{vdv}{\sqrt{2v^3 + 2Ev^2- L^2}}\,.
\end{equation}
A light cone version of (\ref{J13}) with $t (=x^0)\rightarrow \tau:=  \frac{x^0+ x^5}{2}$, i. e.
$$
\vec{e}=  \frac{2}{x^0+ x^5}\sqrt{v}
\begin{pmatrix} 
\cos\theta&\\
\sin\theta&\\
\end{pmatrix}, 
$$
implying $4v= (x^0+ x^5)^{2}(x^{2}_{1} +  x^{2}_{2} +x^{2}_{3} +x^{2}_{4} )$ and obtaining $(x^0- x^5)=:x_-$ 
via integrating  $ \partial_{\varphi}x_- =\partial_{\tau}\vec{x}\partial_{\varphi} \vec{x}$,
 leads to the  extremal {3}-folds

$$
\mathbb{M}_{c,L}= 
\{ \, (x^\mu)\in \mathbb{R}^{1,5} |\,\, (x_0 - x_5)^2 (x^{2}_{1} +  x^{2}_{2} +x^{2}_{3} +x^{2}_{4} +  x^{2}_{0} - x^{2}_{5})= c\;,\; x_1x_4 - x_2x_3 = 0, (*)
\, \},
$$  where the third condition (*) relating the 6 coordinates is obtained by eliminating $v$ and $\theta$, via (24) and (27) (with $8E=-3c$), from $(e_1)^2 = (x_1)^2 + (x_2)^2$ and $(e_2)^2 = (x_3)^2 + (x_4)^2$. 
Various other aspects, including the question of integrability of (\ref{J14}) and  (\ref{J15})   are currently under investigation \cite{H5}.

\subsection*{Acknowledgements}
I would like to thank the organizers of the workshops `Quadrature Domains and Laplacian Growth in Modern Physics', resp. 'Microscopic Description of Singularities', for their kind invitation, the Marie Curie
Training Network ENIGMA and the Swedish Research Council for support,
and P.~Allen, L.~Andersson, M.~Bordemann, J.~Eggers, B.~Gustafsson,
V.~Moncrief, A.~Restuccia, S.~Theisen, M.~Trzetrzelewski, and A.~Zheltukhin for discussions.


\begin{thebibliography}{1}
\bibitem{H6}
 $\mathbb{M}_{2}$ arises as a special case ($\omega=4$, $c=3$) of rotating with angular velocity
$\omega$ shapes given by $r\left( \begin{array}{c}
                        \cos \theta      \\
                        \sin \theta   \\
                              \end{array} \right)$,
\[
r^2(\tilde{\varphi})=\frac{1}{2\omega^2}\left( (c^2+1)+(c^2-1)\cos
k\tilde{\varphi} \right),
\]
\[
\frac{d\theta}{d\tilde{\varphi}}=\frac{1-\omega^2r^2}{1-c^2}\frac{c
\omega}{\omega^2r^2}=\frac{kc}{4}(c^2-1)\frac{1+\cos
k\tilde{\varphi}}{1+\frac{c^2-1}{2}(1+\cos k \tilde{\varphi})},
\]
$k=\frac{2\omega}{c^2-1}$; for $c=(2n+1)=c_n$ an odd integer,
$\omega=2n(n+1)=\omega_n$, $(k=1)$, the non-parametric
description is $\omega_n x=\sqrt{w}\cos\theta=h_{n+1}(w:=\omega^2r^2)$, with
$h_{n+1}$ a polynomial 
(resp. $-\sqrt{w}\cos\left(c\arctan\sqrt{\frac{w-1}{c^2-w}}-\arctan\left(c\sqrt{\frac{w-1}{c^2-w}}\right)\right)$) 
 satisfying
\[
c^2(h^2-w)(w-1)=(h-2w h')^2 (w-c^2).
\]
\end{thebibliography}
\end{document}